\documentclass[preprint,showpacs,aps]{revtex4}

\usepackage{graphicx}
\usepackage{dcolumn}
\usepackage{bm}

\newcommand{\bd}{
\begin{document}}
\newcommand{\ed}{\end{document}}
\newcommand{\beq}{\begin{equation}}
\newcommand{\eeq}{\end{equation}}
\newcommand{\bef}{\begin{figure}}
\newcommand{\enf}{\end{figure}}
\newcommand{\bea}{\begin{eqnarray}}
\newcommand{\eea}{\end{eqnarray}}
\newcommand{\baR}{\begin{array}}
\newcommand{\eaR}{\end{array}}
\newcommand{\bc}{\begin{center}}
\newcommand{\ec}{\end{center}}
\newcommand{\ben}{\begin{enumerate}}
\newcommand{\een}{\end{enumerate}}
\newcommand{\bit}{\begin{itemize}}
\newcommand{\eit}{\end{itemize}}
\newcommand{\su}{\section}
\newcommand{\ssu}{\subsection}
\newcommand{\sssu}{\subsubsection}
\newcommand{\nid}{\noindent}
\newcommand{\nnb}{\nonumber}

\newcommand\obs{A}
\newcommand\cJ{{\sf J}}
\newcommand\be{{\bf e}}
\newcommand\bx{{\bf x}}
\newcommand\bu{{\bf u}}
\newcommand\bfn{{\bf f}}
\newcommand{\bdelta}{\mbox{\boldmath $\delta$}}
\newcommand\bthe{\mbox{{\boldmath $\theta$}}}
\newcommand\bxi{\bm{\xi}}
\newcommand\beps{{\mathbf \epsilon}}
\newcommand\bF{{\bf F}}
\newcommand\bG{{\bf G}}
\newcommand\tbG{{\tilde{\bf G}}}
\newcommand\bX{{\bf X}}
\newcommand\tu{\tilde{u}}
\newcommand\tbu{\tilde{\bu}}
\newcommand\cB{{\cal B}}
\newcommand\cC{{\cal C}}
\newcommand\cX{{\cal X}}
\newcommand\hC{\hat C}
\newcommand\hchi{\hat\chi}
\newcommand\hkappa{\hat\kappa}

\bd

\title{Stable resonances and signal propagation in a chaotic network of coupled units.}
\author{B. Cessac}
\affiliation{Institut Non Lin\'eaire de Nice, 1361 Route des Lucioles, 06560 Valbonne, France}
\author{J.A. Sepulchre}
\affiliation{Institut Non Lin\'eaire de Nice, 1361 Route des Lucioles, 06560 Valbonne, France}

\begin{abstract}
We apply the linear response theory developed in \cite{Ruelle} to analyze how a periodic signal of weak amplitude,
 superimposed upon a chaotic background, is transmitted
in a network of non linearly interacting units.
We numerically compute the complex susceptibility and show the existence of 
specific poles (stable resonances) corresponding to the response to perturbations transverse to the attractor.
Contrary to the poles of correlation functions  they depend on the pair
emitting/receiving units. This dynamic differentiation, induced by non linearities, exhibits 
the different ability that units have to transmit a signal in this network.
\end{abstract}

\maketitle 

\pagebreak

\section{Introduction} 

Currently, there is considerable research activity in network
dynamics. This is clearly motivated by the wide expansion of
communication media (mobile phones, Internet, multimedia, etc.),
but also by the growing interest in network modeling of
biological processes (neural networks, genetic networks,
ecological networks ...). A large part of these studies focuses on
topological properties  of the underlying graph. However, in many cases, the nodes of the networks are
units behaving in a {\it non linear} way. For example, in a
communication network a relay regenerates (amplifies) weak
signals, but it has a finite capacity and saturates if too many
signals arrive simultaneously. A neuron has a non linear response
to an input current, a gene expression is determined by a non linear
function of the regulatory proteins concentration, etc.. These non-linearities might modify
the network abilities in a drastic way. For example, a relay may
have a high graph connectivity (``hub''), but the dynamics
 drives it close to its saturation point, so that it has a weak
reactivity to the changes in the inputs coming from the other
units and a poor capacity to transmit information.  Consequently,
 the information is transmitted via other units, possibly weaker
 links, and, in this regime, these units become temporary ``hubs''
though they may have a low graph connectivity, while the main hub
is decongested. In biological networks similar effects may arise.
For example, the capacity of a neuron to transmit a specific excitation
strongly depends on its state, determined itself by the overall currents
coming from afferent neurons.  

This  suggests us that the mere study of the
graph topological structure of a network with non linear units 
is not sufficient to capture the full dynamical behavior.
However, there are relatively few studies which analyze the joint effect of
topology of the network and  non-linearity.
 Nevertheless, these networks are dynamical systems
with a large number of degrees of freedom, and so dynamical
systems theory and statistical mechanics provide a powerful
framework to state problems in a well-defined way and to propose
solutions. \\

In this paper, we analyze the following situation. 
We consider a  network composed by a set of $N$  units
receiving and transmitting signals. At each time step $t$ the unit $i$ receives a bench of signals coming 
from each unit connected to it, and it emits, at time $t+1$, a signal which is a sigmoid function of the 
global input [see eq. (\ref{DNu})]. In the model studied below, the global dynamics has generically a chaotic attractor,
provided
that the non linearity of the transfer function is sufficiently large [see Section \ref{Model}].
In spite of the presence of chaos it is possible to analyze how a periodic signal of weak amplitude,
 superimposed upon a chaotic background, is transmitted
in the network. However, as discussed above,
this analysis requires the consideration of the network structure \textit{as well as} nonlinear effects.
 
The main tool we use for this investigation is the  linear response theory developed by D. Ruelle \cite{Ruelle}
for hyperbolic dynamical systems (e.g. dissipative systems with a chaotic attractor)
in a non equilibrium steady state. This theory allows us to 
 compute explicitly the variation of the average value of a generic
observable, induced by a time dependent signal of weak amplitude.
Indeed, provided that the amplitude of the signal is sufficiently small (but \textit{finite}),
this variation is a linear function of the signal and a linear response operator
is explicitly given in terms of the dynamic evolution. In our case, this operator has a simple
expression (see eq. (\ref{Chi})). 
The effects of a periodic signal emitted by a unit on a receiving unit
is characterized by the Fourier transform of the linear response, 
called \textit{susceptibility} in the sequel (see section \ref{CSus}).
This gives us a frequency response curve (see Fig. \ref{compare_pics_sus}) exhibiting resonances peaks.
 These resonances corresponds to complex poles for the analytic
continuation of the susceptibility in the complex plane. They have a nice interpretation in Ruelle theory.

Indeed, in this theory, the linear response operator is
 the sum of two contributions. There is a regular term,
corresponding to the response to  perturbations ``parallel'' to the attractor 
(more precisely locally projected along the unstable
 manifold). This term is actually a correlation
function \cite{RuelleCMP} and, consequently, it obeys classical relations such as the Fluctuation-dissipation theorem.
 The poles of its
Fourier transform are called  Ruelle-Pollicott resonances \cite{RP} or ``unstable'' poles.
They give the rate of mixing of the chaotic system
or equivalently, the relaxation rate to equilibrium for a perturbation ``on''  the
attractor. These poles are independent of the observable. Therefore, in our case,
they are independent of the pair emitting/receiving unit  (see Fig \ref{Poles_Ruelle_Cij4096}). When focusing on the response
to real frequency one observes therefore resonance peaks common to all pair of units,
and these peaks are also present in the  Fourier spectrum of the corresponding correlation function.

The second term  corresponds to the response to perturbations
locally projected along stable manifolds, namely \textit{transverse} to the attractor. 
Therefore, it exists only in the dissipative case. It does
not obey fluctuation-dissipation theorem and has drastically different properties
than the first term. In particular its poles (``stable'' poles) are expected to be distinct from
the unstable poles. 
In this paper, we indeed exhibit such stable poles. To the
best of our knowledge, this is the first example where
these poles are explicitly exhibited, though their existence was theoretically proved.
 Moreover, we show numerically that the stable poles depend on the pair 
emitting/receiving unit (see Fig. \ref{Compare_poles_sus}). When focusing on the response
to real frequency one observes therefore \textit{specific} resonances peaks 
(see Fig. \ref{compare_pics_sus}). This shows that  
a unit receiving a periodic signal emitted from another unit may respond in
a specific way to this signal, the amplitude depending both on the signal
frequency and \textit{on the emitting unit}. Note that according to the discussion
above this effect cannot be observed by studying correlation functions. \\

The paper is organized as follows. In section \ref{Model} we introduce the model
and discuss its properties. The section \ref{RepLin} recall briefly
the main results of Ruelle linear response theory used in this paper.
An explicit computation of the linear response is performed. It shows
the explicit contributions of the network topology and of the non linearity in a signal propagation.
In section \ref{CSus} we compute numerically the frequency response curve and discuss
the different resonance peaks.  The poles of the complex susceptibility for a few
pairs of units are computed and compared in the section \ref{Poles}. Our main conclusions
are then drawn. 

\su{Model} \label{Model}

Consider the following 
dynamical system, originally proposed
in the context of Neural Networks [see \cite{IJBC,PD,JP} and references therein].
The output signal  is a function of the weighted sum 
of the signals arriving at $i$ at time $t$
and is given by:
\begin{equation}
u_i(t+1)= \sum_{j=1}^{N}  J_{ij} f(u_j(t))
\label{eq:nn}
\end{equation}
The weights $J_{ij}$'s may be  positive (excitatory), negative (inhibitory) or
zero (no direct link between $j$ and $i$). They are  in general non symmetric
($J_{ij} \neq J_{ji}$). Thus, the matrix of weights, $\cJ$,
 defines an oriented graph such that there is a link from $j$ to $i$ if and only
if $J_{ij} \neq 0$. The global dynamics
can also be written as:
\beq\label{DNu}
\bu(t+1)=\bG\left[\bu(t)\right]=\cJ\, \bfn (\bu(t)),
\eeq
\nid where  $\bu(t) = \left\{u_i(t)\right\}_{i=1}^{N}$ and
where we used the notation $\bfn (\bu(t)) = \left\{f(u_i(t))\right\}_{i=1}^{N}$.
Consider now the case where the nonlinear transfer function  $f$ is a
 sigmoid, [e.g. $f(x)=\tanh(gx)$], where
the parameter $g$ controls the non linearity. In terms of input/output ratio,
a unit amplifies weak signals (if $g>1$), but with a limited capacity: $f$  ``saturates'' if the local field 
 is too strong, and the variations of the output signal are all the weaker as the local field is big.
Thus, the capacity of $i$ to retransmit a signal emitted from $k$ \textit{does not only depend
on the weight $J_{ik}$ but also on the state of saturation of $i$ when it receives the signal coming
from $k$}. Note also that the Jacobian matrix $D\bG(\bu)$ writes $D\bG_{ij}(\bu)=J_{ij}f'(u_j)$ where
$f'$ is the derivative of $f$. Therefore, the volume variation is proportional to $\prod_{i=1}^N f'(u_i)$.
Therefore, in this model, the dynamical contraction is closely related to the saturation of the sigmoid transfer
function.

In order to emphasize the effects of the nonlinearity and minimize the effect of the network topology, one may
assume that  the network is fully connected  and that
the $J_{ij}$'s are drawn randomly with respect to some smooth distribution (uniform, Gaussian,
etc \dots). As an example, one may fix the average value $[J_{ij}]=0$ and
the  variance $[J_{ij}^2]=\frac{1}{N}$ (to ensure the correct normalization of the local field with
the size $N$). This example is interesting because the system (\ref{DNu})
 exhibits a wide variety of dynamical regimes (static, periodic, quasi periodic, chaotic). More precisely,
it has been shown in \cite{PD} that it generically exhibits a transition to chaos by quasi periodicity
 when $g$ increases. Note that the same transition occurs if the network
 is sparse \cite{IJBC} with $K>2$ neighbors ($K$ can be random) chosen at random, provided
the variance of the $J_{ij}$'s scales like $\frac{1}{K}$. However, we  do not address this case in this paper
since we want to minimize the effect of the network structure. Note also that this type of transfer
function allows dynamical regimes where several attractors coexist. It has been
indeed shown in \cite{PD,JP} that, adding a  threshold $\theta$ to the local field, there exists a region in the parameter
space $g,\theta$ where two attractors coexist. This region can be analytically computed. However,
in the present paper, the parameters are located in a region where  there is only one attractor and
all initial conditions converge to this attractor. \\

Let us now assume that the non linearity is large enough so that the global dynamics has a chaotic attractor
(with all Lyapunov exponents bounded away from zero and at least one positive Lyapunov exponent).

 We  now add a signal  of small amplitude $\bxi(t)$ to the output of some units. Then 
the evolution of the perturbed system, denoted by $\tbu$, 
 is  given by :
\beq \label{pert}
\tbu(t+1)=\bG\left[\tbu(t)\right]+\bxi(t)=\tbG\left[\tbu(t)\right].
\eeq 
Note that the formalism introduced below accommodates the generalization 
where $\bxi(t)$ depends also  on  $\bu(t)$, but  we do not consider this case here. 

We want to investigate the capacity of the network to transmit  signal $\bxi(t)$  superimposed upon the chaotic background.
This is a complex problem since after a few time steps the total signal arriving
at time $t$ at $k$ includes the sum of contributions corresponding to different paths followed by $\bxi$, 
with different time delays. Moreover, along
a path the signal can
be damped if $f$ saturates  ($f' < 1$), or amplified ($f' > 1$).
Finally, the dynamics being chaotic, after a sufficiently long time the signal is distorted by the nonlinearities
and scrambled by mixing.

To tackle this problem we  analyze how the difference $\tbu(t)-\bu(t)$ between the perturbed and unperturbed 
dynamics behaves  \textit{on average}  as a function of  $\bxi(t)$. 
When $\bxi(t)$ is small enough, and in spite of the initial condition sensitivity intrinsic to chaotic systems,
it can be shown that this difference is a \textit{linear} functional of $\bxi(t)$.
This is the content of the {\em linear response theory} developed by D. Ruelle \cite{Ruelle} for chaotic and dissipative
\footnote{Dissipative means here that the phase space volume is contracted by the dynamic evolution.}
system. Some aspects of this theory are briefly recalled in the  next Section.

\section{Linear response theory.}  \label{RepLin}

The unperturbed dynamical system (\ref{DNu}) has a strange (chaotic) attractor for sufficiently large $g$.
Usually, strange attractors  
carry a natural probability measure called the Sinai-Ruelle-Bowen (SRB) measure \cite{SRB}. 
If one prepares the system  (\ref{DNu})
with an initial macrostate $\mu$ having a uniform density ({\em i.e.}  $\mu (d\bu) = d\bu$), 
corresponding to selecting \textit{typical} initial conditions, then,
 provided that the limit exists, the SRB measure is the asymptotic macrostate $\rho=\lim_{t \to +\infty} \bG^{t} \mu$ 
where $\bG^{t}\mu$ is the image of $\mu$ under the $t$-th iterate of $\bG$. 
The SRB measure has several remarkable features which make it ``natural''
\footnote{Sinai, Ruelle and Bowen have indeed shown that the SRB measure is a Gibbs like measure. 
Moreover it maximizes some version 
of a free energy (topological pressure) : it has therefore the characteristics of an equilibrium state.
A crucial property for the present work is that a SRB measure
 has a density along the unstable manifolds,
 but it is singular in the directions transverse to the attractor.  
This feature is at the origin of the distinction between unstable 
and stable poles of the susceptibility.}.
One of its most important property for practical purpose is the following:
If $A$ is some observable (a smooth function of $\bu$), its average with respect to $\rho$,
\begin{equation}
< \obs >=  \int \obs(\bu) \, \rho(d\bu) 
\label{eq:SRBmean}
\end{equation}
 is equal to the time average along 
\textit{typical trajectories}.  This means that ``ensemble average'' and time average are equivalently
for typical trajectories. This is especially useful
 for numerical computations (see next Section).  \\

Applying  a time dependent perturbation $\bxi(t)$ to the system induces 
time dependent changes in the statistical averages. More precisely, the natural extension of the SRB measure defined above is  
 a  \textit{time dependent} SRB measure $\rho_t$. It is given by 
the (weak) limit $\lim_{s \to +\infty} \tbG^t  \dots  \tbG^{t-s} \mu$.   
The  corresponding average will be denoted  by $< \ >_t$. 

It has been  established in \cite{Ruelle}
 that \textit{a linear response theory exists}
for \textit{uniformly  hyperbolic diffeomorphism\footnote{We only know
that the system (\ref{DNu}) is {\em weakly} hyperbolic, i.e. all the Lyapunov exponents are bounded away from zero. 
Nevertheless we will adopt the point of view defended
in \cite{Ruelle}. If there is a linear response theory for our system, it is necessarily
of the form eqs~(\ref{eq:delta_u})-(\ref{Chi}), since there are no reasonable alternative. What could happen
is that the sum diverges leading to an infinite response. On numerical grounds,
one has to check that the time average used to compute the ergodic average [see eq.
(\ref{susth})] does not increase with the sample length.}}.
In our framework, this means that, provided that $\bxi(t)$ is sufficiently small,
and for any smooth observable $\obs$, the  variation $< \obs >_t-< \obs >$ is \textit{proportional} to $\bxi(t)$ up to small 
non linear corrections.
In other words,  $\rho_t $ is \textit{differentiable}  with respect to the perturbation.
The derivative is called the  \textit{linear} response.\\

The theory developed by Ruelle allows one to compute the linear response,
 for general perturbations depending both on time $t$
and state $\bu$, and for a general observable $\obs$.  In our context, however, where the considered observables are simply the variables of systems~(\ref{DNu})--(\ref{pert}), the linear response has a simple form, which can be written as:
\begin{equation}
<  \tbu>_t - <  \bu > = \sum_{\tau = -\infty }^{\infty}   \chi (\tau) \, \bxi(t-\tau-1)
\label{eq:delta_u}
\end{equation}
where  $\chi(\tau)$ represents the averaged Jacobian matrix: 
\beq\label{Chi}
\chi(\tau) = < D\bG^{\tau} (\bu)> ,
\eeq
for $\tau \geq 0$.  Otherwise $\chi(\tau) = 0$  (which is consistent with the requirement of  causality). 

A remarkable consequence of Ruelle theory  is that $\chi(\tau)$ is a bounded function for all $\tau \geq 0$.
In particular, it does not diverge exponentially fast, despite the presence of 
a positive Lyapunov exponent. As discussed below, this is essentially a consequence of exponential mixing.

In what concerns network dynamics, equation~(\ref{eq:delta_u})  is  interpreted as giving  the average response of unit $i$
 of the system  when the network  is submitted to weak signal $\bxi(t)$.    
In particular it is seen that if  only one  unit $j$ is perturbed  at time $t=-1$ by a kick of amplitude $\epsilon$   
[that is $\bxi (t) =  \epsilon \be_j \delta (t+1)$ with  the Kroenecker symbol $\delta$ and  the $j$-th unit vector $\be_j$],  
then  $\epsilon \, \chi_{ij} (t) $ gives precisely the mean \textit{response} of unit $i$ at time $t$.   
This  suggests to define the {\em susceptibility}  of the network  as the Fourier transform of $\chi_{ij}(t)$, namely:
\begin{equation}
\hat{\chi}(\omega) = \sum_{t=-\infty}^{\infty}  \chi(t) e^{i \omega t}
\label{eq:susceptibility}
\end{equation}
This matrix function will be numerically computed and studied in the next Section.   
We conclude the present Section by analyzing further the structure of $\chi_{ij}(\tau)$ 
in the case of dynamical system~(\ref{eq:nn}).  Here one can decompose $\chi_{ij}(\tau)$ as :
\beq\label{chiij}
\chi_{ij}(\tau)=
\sum_{\gamma_{ij}(\tau)}
        \prod_{l=1}^{\tau}J_{k_l k_{l-1}}
\left< \prod_{l=1}^{\tau}f'(u_{k_{l-1}}(l-1))\right>,
\eeq
\nid  The sum holds on each possible paths
$\gamma_{ij}(\tau)$, of length $\tau$, connecting the
unit $k_0=j$ to the unit $k_\tau=i$, in $\tau$ steps.
One remarks that each
path is weighted by the product of a \textit{topological} contribution
depending only on the weight $J_{ij}$ and
a \textit{dynamical} contribution. Since, in the kind of systems we consider, functions $f$ are sigmoids, 
the weight of a path $\gamma_{ij}(\tau)$ depends crucially on
the state of saturation of the units $k_0, \dots, k_{\tau-1}$ at times $0, \dots, \tau-1$.
Especially, if $f'(u_{k_{l-1}}(l-1))>1$ a signal is amplified while it is damped if
$f'(u_{k_{l-1}}(l-1)) < 1$. Thus, though a signal has many possibilities for going from $j$ to $i$ in $\tau$ time steps,
some paths may be ``better'' than some others, in the sense
that their contribution to $\chi_{ij}(\tau)$ is higher.  Therefore eq.~(\ref{chiij})  underlines a key point discussed in the introduction.  The analysis of signal transmission in a coupled network of dynamical units requires to consider both the topology of the interaction graph {\em and}  the nonlinear dynamical regime of the system.

\section{Complex susceptibility.}\label{CSus}

One can decompose the response function (\ref{Chi})  into two terms.
The first one is obtained by locally projecting the Jacobian matrix  on the unstable directions of the tangent space.  This term will be named  the``unstable" response function.
It corresponds to linear response of the system to  perturbations locally parallel to the local unstable manifold (roughly speaking perturbations ``on'' the attractor).
One can show that the linear response associated with this type of perturbation is in fact  a  correlation function, 
as found in  standard fluctuation-dissipation theorems~\cite{Ruelle}. 
Hence, as usual for correlation functions of a chaotic system,  \textit{it decays exponentially} (because of mixing) and the decay rates are associated with the poles of its Fourier transform.  More precisely, 
these exponential decay rates correspond to the imaginary part of the complex poles of the unstable part of the 
susceptibility~(\ref{chiij}).  Thus they will be called  ``unstable'' poles.
More generally, it can be shown that these poles  are also the eigenvalues of the operator governing the time-evolution of the probability densities (which we denoted above as $\bG^{t} \mu$), the so-called Perron-Frobenius
operator~\cite{RP}.  Therefore, these poles, whose signatures are visible in the peaks of the modulus of the correlation functions, do not depend on the observable, though some residues may accidentally
vanish for a given observable.   

The second term \footnote{Note that a linear response theory has also been proposed in \cite{Vulpiani}.
However, it requires  the invariant measure to have a density. This is only true for the conditional
measure along unstable manifolds. As a matter of fact, this theory does not contain the stable term.}
 is obtained by locally projecting the Jacobian matrix  on the stable directions of the tangent space. 
It corresponds to the response to perturbations  locally parallel to the local stable manifold 
(namely transverse to the attractor).  Therefore, it is \textit{exponentially damped} by the dynamical contraction. 
[Note that, according to the specific
form of the Jacobian matrix, this contraction is, in our case, mainly due to the saturation of the sigmoid
transfer function].
The corresponding exponential decay rates are given by the complex poles (``stable'' poles) of
the stable part of the complex susceptibility. But here the poles depend a priori on the observable.
One can easily figures this out if one decomposes the stable tangent space of a point in the
orthogonal basis of Oseledec modes (directions associated to each of the negative Lyapunov
exponent). The projection of the $i$-th canonical basis vector on the $k$-th Oseledec mode 
depends on $i$ and $k$. This
dependence persists even if one takes an average along the trajectory, as in (\ref{Chi}). 

Hence, both stable and unstable terms are exponentially damped, ensuring the convergence of the
series (\ref{eq:delta_u}), but for completely different reasons. Moreover, the stable and unstable 
part of the linear response have drastically different properties. As a matter of fact, the stable part
\textit{is not a correlation function and it does not obey the fluctuation-dissipation
theorem}. In particular, the unstable poles and stable poles
are expected to be distinct. In this
paper, we give for the first time an evidence of this theoretically predicted effect.
Moreover, we show that the stable poles indeed allow to distinguish the units in their
capacity to transmit a signal. \\

For this we first numerically compute the susceptibility~(\ref{eq:susceptibility}) for real
values of $\omega$.  The computation is based on the following remark.
Let us consider perturbations
  $\bxi^{(1)}(t)= \epsilon \be_j \cos(\omega t)$
and  $\bxi^{(2)}(t)= - \epsilon \be_j \sin(\omega t)$ 
and let $\bu^{(1)},\bu^{(2)}$ denote the variables of the corresponding perturbed systems:
\beq
\label{pert_k}
\bu^{(k)}(t+1)=\bG\left[\bu^{(k)}(t)\right]+\bxi^{(k)}(t)  \quad \quad (k=1,2)
\eeq
Then it follows  from (\ref{eq:delta_u}) that :
\bea
(<  u_i^{(1)}>_t - <  u_i >) + i ( <  u_i^{(2)}>_t - <  u_i > ) & = & \epsilon
\sum_{\tau}   \chi_{ij}  (\tau) \, e^{-i\omega(t-\tau-1)} \nonumber \\
& =&  \epsilon \hat{\chi}_{ij}(\omega) \, e^{-i\omega(t-1)} 
\label{combined}
\eea
Note that the time dependent average response
to periodic perturbation is therefore periodic.  
The linear response at time $t$ is an infinite sum corresponding
to contributions of time delayed signals following different paths. Since the signal is sinusoidal
the terms in this sum may interfere in a constructive way [but exponential damping prevent
the series to diverge, ensuring the existence of a linear response].

Since $\hat{\chi}_{ij}(\omega)$ is independent of $t$, then it is equal (for $\omega \neq 0$) to the time average 
\beq
\hat{\chi}_{ij}(\omega) = \lim_{T\to\infty}{1\over T\epsilon}\sum_{t=0}^T e^{i\omega(t-1)}\,
[<  u_i^{(1)}>_t  + i  <  u_i^{(2)}>_t ]
 \eeq
The time-dependent averages  $ <  u_i^{(k)}>_t $
 involve an average over initial conditions in the distant past.
  One can reasonably assume that the above average over $t$ makes the average over the initial conditions unnecessary. 
 Then one may write :
\beq\label{susth}
\hat{\chi}_{ij}(\omega) =\lim_{T\to\infty}{1\over T\epsilon}\sum_{t=0}^T e^{i\omega(t-1)}\,
[  u_i^{(1)}(t)  + i  u_i^{(2)}(t )] 
\eeq
where the $u_i^{(k)}(t)$  $(k=1,2)$ are obtained by iterating maps~(\ref{pert_k}).
 This provides 
a straightforward way to compute
the susceptibility, where most of the computing time 
 goes into computing the orbits $\bu^{(k)}(t)$. \\

As an example, we performed a numerical computation of the dynamical system (\ref{DNu})
where we take a fixed realization of  $J_{ij}$'s, with  $N=8$ units.
There is a quasi-periodic transition to chaos as $g$ increases.
The system is studied for  $g=3.5$ corresponding to  a positive Lyapunov exponent $\lambda_1=0.158$, while the second
one is $\lambda_2=-0.183$. The system is therefore weakly hyperbolic (all Lyapunov exponents bounded away
from $0$).

The function $\hchi(\omega)$, the Fourier transform of the matrix (\ref{chiij}),
has been computed with a resolution $\delta \omega=\frac{\pi}{2048} \approx 1.53 \times 10^{-3}$. The average
is done with $26214400$ samples. 
We did several runs where we varied the length $T$ of the time average in (\ref{susth}). We checked that the global structure is the same.
In particular the amplitude of the susceptibility $|\hchi(\omega)|$ \textit{does not depend on $T$} 
(see note [12]). Also  the fluctuations decrease 
like $\frac{1}{\sqrt{T}}$ according to the central limit theorem. 
 
\begin{figure}
\begin{center}
\includegraphics[width=12cm,clip=false]{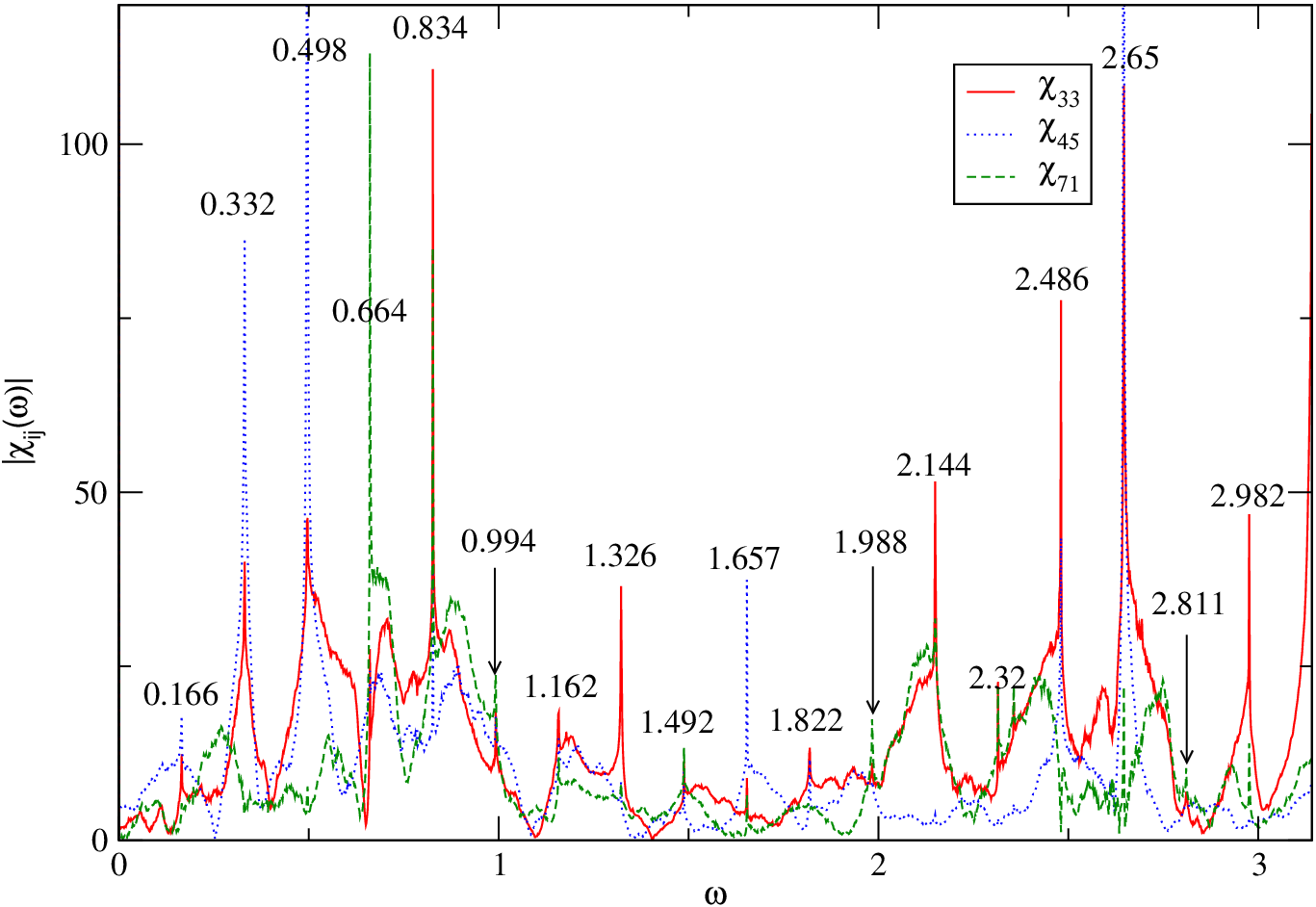}
\caption{\label{compare_pics_sus} \footnotesize (Color on line)Modulus of the susceptibilities 
$\hchi_{33}$ (red),$\hchi_{45}$ (blue),$\hchi_{71}$ (green).}
\end{center}
\end{figure}

In Fig. \ref{compare_pics_sus} we have plotted the modulus of the susceptibilities 
$\hchi_{33},\hchi_{45},\hchi_{71}$. Comparing these functions, 
one remarks  that there are thin peaks essentially 
 located at the same frequencies, with different heights. Moreover, these frequencies are harmonics
of a fundamental frequency ($\omega_0 \sim 0.166$). 
This is expected from the frequency locking in the quasi-periodic transition preceding  chaos. 
Some of these frequencies
are also present in the Fourier spectrum of the correlation functions
 but with a smaller amplitude and some peaks are indistinguishable from the background.
Instead, all harmonic peaks are revealed in the susceptibility spectrum.

But we also note that for many peaks, the \textit{width} 
varies strongly from a pair $ij$ to another. 
This means that the \textit{resonance strength}
depends on which unit is excited and which unit responds. In particular, some peaks
are very thin, corresponding to an accurate resonance while some others are broad. 
In terms of poles, this means that the imaginary part are distinct and consequently
the corresponding poles are different [see next section].
Finally there are \textit{additional} peaks  strongly dependent on the pair $ij$. 

Thus, a simple glance to Fig. \ref{compare_pics_sus} tells us
that the frequency response of a unit $i$ to the excitation emitted
by a unit $j$ strongly depends on the pair $i,j$. As discussed above,
and numerically shown below, this difference comes from the stable
part of the linear response. Consequently, the specificity of the response
is revealed only if one consider perturbations \textit{transverse} to the attractor.
[Note that, generically, the signal is a perturbation having local projections
both on local stable and unstable spaces.]

%
%
%
%
%
%
%
%
%
\section{Unstable and stable poles.} \label{Poles}

 Resonances correspond to poles
in the complex plane. As a matter of fact, the position of the maximum of the peak corresponds to the real part of the pole, 
its width is related to its imaginary part, and the value of the maximum is related
to the residue. From this observation, we
developed an algorithm to estimate the residue width and locations of the poles. Let $\omega_0=\omega_r + i\omega_i$
be a simple pole of $\hchi$ and $A$ its residue. If one multiplies  $\hchi$ 
by a phase factor $e^{i\psi}$ then the real and imaginary part rotate continuously, without
changing the modulus. If the pole is close enough to the real axis then there exists a phase $\psi$ 
such that, on the real axis, the real part has a characteristic Lorentzian shape symmetric
with respect to $\omega_r$ while the imaginary part is antisymmetric. Then a nonlinear 
curve fitting allows us to determine $A,\omega_r,\omega_i$. Once a local analysis has roughly determined
the poles, a global nonlinear fit (Levenberg-Marquardt \cite{NR}) allows us to localize
the poles with a better accuracy.

\begin{figure}
\begin{center}
\includegraphics[width=12cm,clip=false]{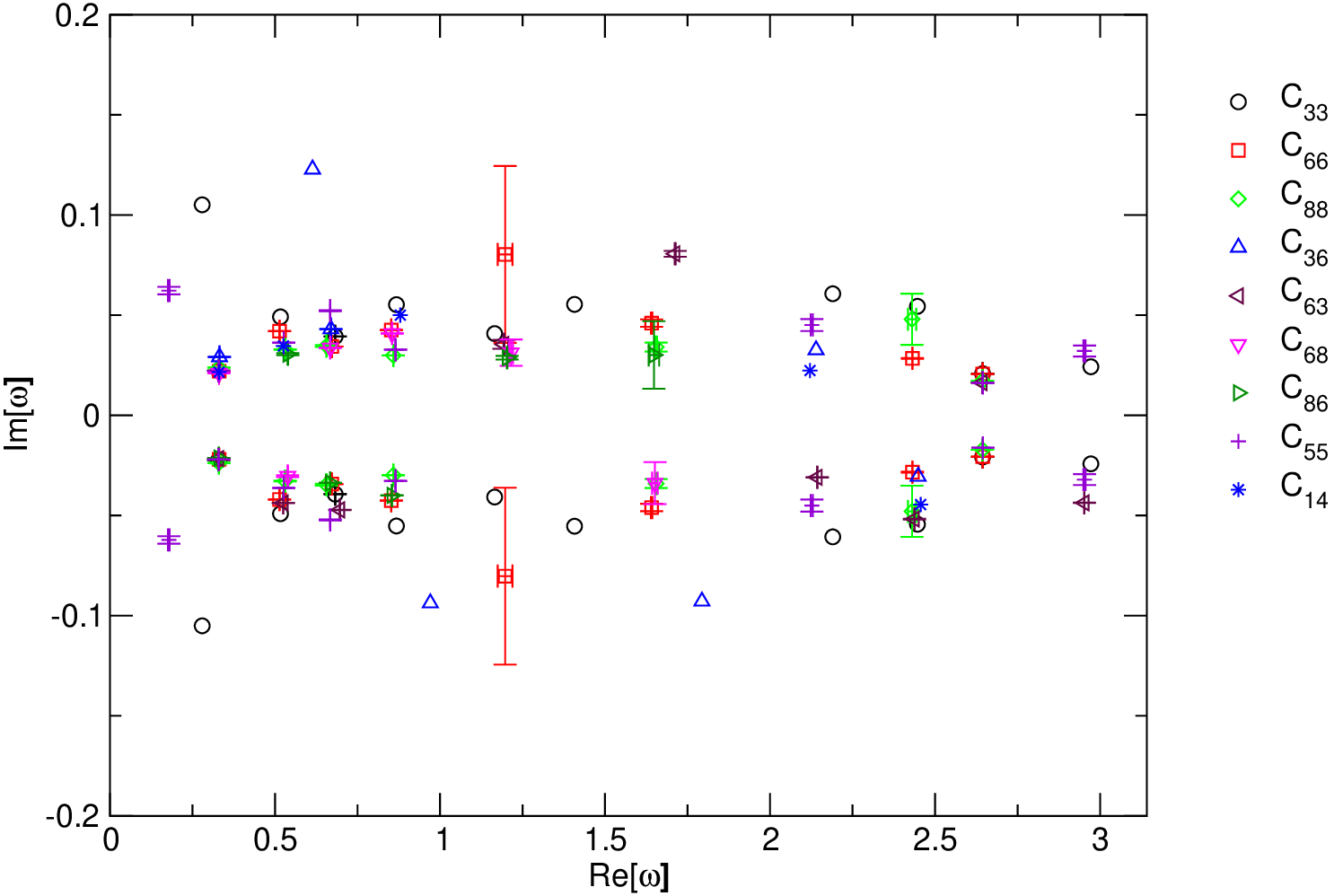}
\caption{\label{Poles_Ruelle_Cij4096}\footnotesize (Color on line) Poles of several correlation functions.}
\end{center}
\end{figure}

\begin{figure}
\includegraphics[width=14cm,clip=false]{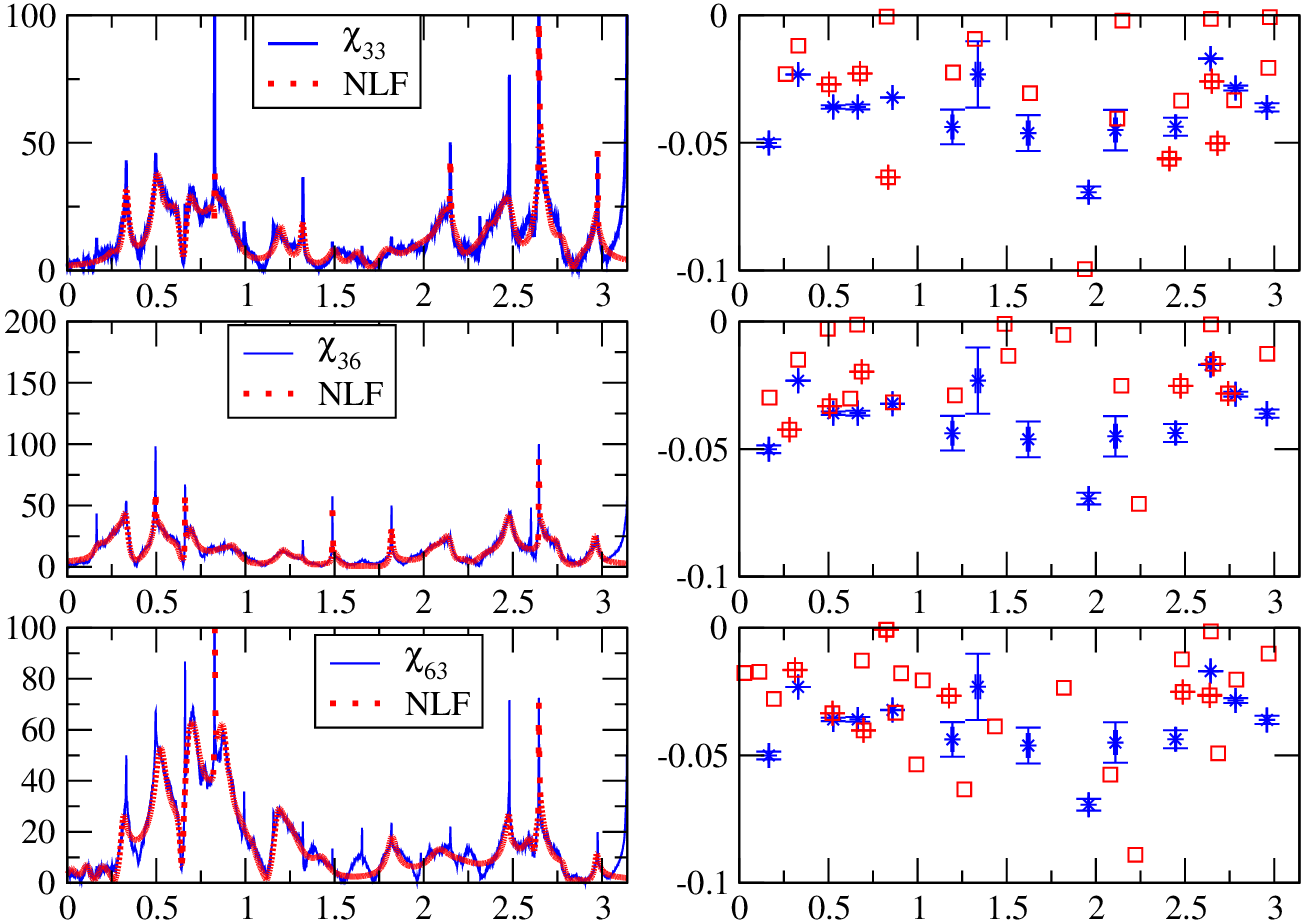}
\caption{\label{Compare_poles_sus}(Color on line) Left column:  Susceptibilities $\hchi_{33},\hchi_{36},\hchi_{63}$  and
reconstruction by the nonlinear fitting algorithm (NLF) used to compute the poles.
 Right column: Poles of susceptibility (red squares) and poles of correlations (represented
by a blue star).}
\end{figure}

In fig.
 \ref{Poles_Ruelle_Cij4096} 
we have plotted the real and imaginary part of the poles of several correlation functions. One notices that 
all pair of units have poles at the same value of $\omega$, within the error bars. We have also 
plotted in Fig. \ref{Compare_poles_sus} the modulus of the
susceptibilities $\hchi_{33}$,$\hchi_{36}$,$\hchi_{63}$ (left column)
and the corresponding poles (right column) with the poles of the correlation functions.
As expected from Fig. \ref{compare_pics_sus} we observe common poles (unstable poles) but also \textit{distinct poles (stable poles)
that, moreover, strongly  depend on the pair receiving/emitting unit}.
%
%
%
%
%
%
%
%
%
%
%
%
%
%

Finally, note that some poles are very close to the real axis. Since their imaginary part is
related to the coherence time of the response to a kick, this tells us that the response
to a pulse may subsist on quite a bit long times, though the underlying dynamics
is chaotic. [Recall however that the linear response measures variations of \textit{average}
value of observables]. This intriguing and exciting aspect will be developed elsewhere.

\section{Conclusion.} This paper gives an example of network dynamics
where  the nonlinearity induces particularly prominent effects
that cannot be anticipated by the mere analysis of the graph topology.
In particular we exhibit a dynamic differentiation in the capacity that a unit has to transmit information.
We also argue on theoretical grounds, and numerically show (see Fig. \ref{Poles_Ruelle_Cij4096})
that  the dynamics differentiation is not revealed by correlation functions. It is purely
an effect of the dynamics transverse to the chaotic attractor that must be handled
with the proper tool. We show that the linear response gives quite a bit more information
than the correlation function, provided that its computation takes into account
the singularity of the SRB measure transversally to the attractor. This is the case
with Ruelle linear response theory and this 
 opens the perspective for developing an extension
of statistical mechanics for the analysis of networks dynamics with nonlinear units.

\begin{acknowledgments}
This paper greatly benefited from the help of David Ruelle.
We deeply acknowledge him for helpful ideas, suggestions and comments.
 We warmly thanks J.L. Meunier for useful remarks. B.C. is grateful to Ph.
Blanchard for all his advices and support.
\end{acknowledgments}

\ed